# Auto-Encoder Optimized PAM IM/DD Transceivers for Amplified Fiber Links

Amir Omidi Member, IEEE, Mai Banawan Member, IEEE, Erwan Weckenmann Member, IEEE, Benoît Paquin Member, IEEE, Alireza Geravand Member, IEEE, Zibo Zheng Member, IEEE, Wei Shi senior, IEEE, Ming Zeng Member, IEEE, and Leslie A. Rusch Fellow, IEEE, Fellow, Optica

Abstract—The use of semiconductor amplifier in integrated transceivers increases sensitivity, but changes the noise statistics in pulse amplitude modulation (PAM) intensity modulation with direct detection. Using a straight-forward, mixed noise model, we optimize constellations for these systems with an autoencoder-based neural network (NN). We improve required signal-to-noise ratio (SNR) by 4 dB for amplified spontaneous emission (ASE)-limited PAM4 and PAM8, without increasing system complexity. Performance can also be improved in O-band wavelength division multiplexing systems with semiconductor optical amplification and chromatic dispersion (CD). At 53 Gbaud, our simulations show we can extend the reach of PAM4 by 4 to 8 km when combining an optimized constellation with a NN decoder. We present an experimental validation of 4 dB improvement of an ASE-limited PAM4 back-to-back transmission at 60 Gbaud using an optimized constellation and a NN decoder.

## I. Introduction

Optical communication with intensity modulation and direct detection (IM/DD) systems are simple and cost-effective. Throughput can be increased with the use of pulse amplitude modulation (PAM), but this requires an increase in signal quality. For example, in [1], a semiconductor optical amplifier (SOA) at the transmitter was used to enable PAM for data center applications. In [2], a SOA was used at the receiver to extend the PON link budget. The SOA will introduce ASE, especially at higher gain levels. We examine the use of optimized PAM constellations for amplified links.

The square law detection of ASE in IM/DD leads to non-Gaussian noise statistics. Unequally spaced PAM has been proposed to improve the receiver sensitivity of amplified PAM IM/DD systems [3], [4]. Based on an approximate analytical model for non-Gaussian and signal-dependent noise, unevenly spaced PAM levels and decision thresholds were found in [3] via an iterative algorithm. They considered only a memoryless channel. An experimental feedback method was adopted in [4] to optimize the intensity levels and decision thresholds of amplified PAM4 at 1310 nm where the chromatic dispersion (CD) is zero. We exploit autoencoder (AE) structures to find optimized unequal PAM levels both for memoryless channels and for channels with memory, such as dispersion or inter-symbol interference (ISI).

We investigate two channel scenarios. For both cases the AE encoder is used only to identify unequal PAM levels; the encoder computation is done once. For the memoryless channel, the AE decoder portion is used only to identify threshold levels; the decoder computation is performed once. For a channel with memory, the AE decoder is trained as a detector instead of thresholds. We keep the complexity of the decoder as low as possible to keep our solution simple and cost-effective.

Machine learning has been used to address several challenges in IM/DD systems, including non-linear pre-distortion in [5] and signal equalization in [6]–[16]. The end-to-end learning framework was also adopted by [17]–[20] for IM/DD systems. In [20], they focused on resilience to dispersion variations using end-to-end learning for an IM/DD system. Their model used hundreds of nodes, whereas ours uses 15. Fiber dispersion in IM/DD was addressed with an AE in [17], [18], but with recurrent neural networks (RNNs) with much greater complexity than what we consider here.

Although previous research used AEs for diverse impairments in IM/DD systems, to the best of our knowledge, none of these studies has specifically addressed mitigation of ASE noise in amplified links with CD for PAM. Our contributions are

- Constellations and thresholds for PAM4 and PAM8 for a memoryless channel with an arbitrary weighted mix of ASE and thermal noise;
  - improvement of 4 dB in required SNR at high ASE with no added complexity;
- Constellations and AE decoders for PAM4 in channels with ASE and memory due to CD;
  - reach improvement of 4-8 km at complexity increase of less than an order of magnitude compared to linear filtering;
- Detailed analysis of experimental validation of improvement for high ASE (presented in part at [21]);
  - experimental 4 dB improvement in optical signal-to-noise ratio (OSNR) at no added complexity.

The paper is organized as follows. Section II briefly presents end-to-end learning for PAM transceiver optimization, and

A. Omidi, E Weckenmann, B. Paquin, A. Geravand, Z. Zheng, W. Shi, M. Zeng, and L. A. Rusch are with the Centre for Optics, Photonics and Lasers, Université Laval, Quebec, QC, G1V 0A6, Canada (e-mails: (amir.omidi.1, erwan.weckenmann.1, benoit.paquin.3, alireza.geravand.1, zibo.zheng.1)@ulaval.ca; (wei.shi, ming.zeng, leslie.rusch)@gel.ulaval.ca). Mai Banawan is now with Alexandria University, Electrical Engineering Department, Alexandria 21544, Egypt (email: Mai.a.f.banawan@alexu.edu.eg).

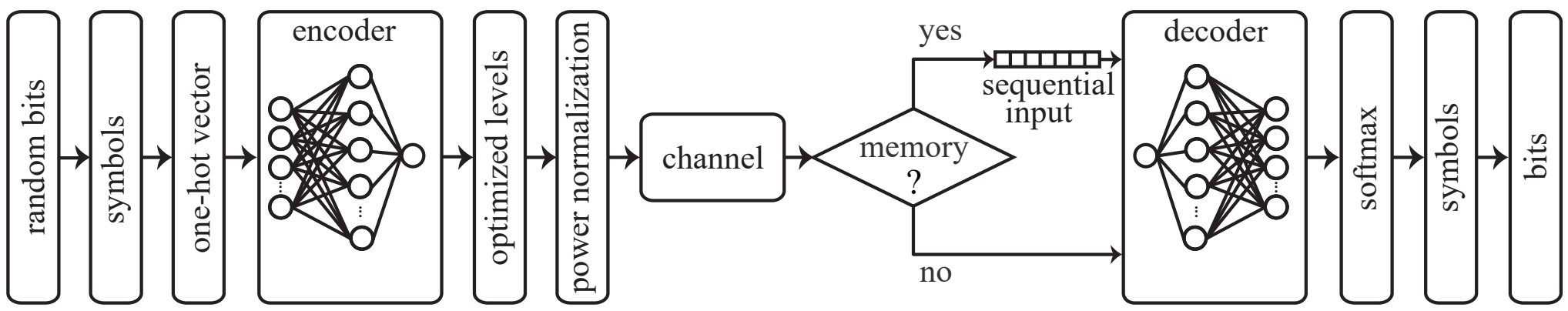

Fig. 1. Block diagram an autoencoder (AE) for communications systems

noise statistics in a memoryless amplified IM/DD channel (detailed development in Appendix A). In Section III, we present performance improvement for a memoryless channel (no CD). Appendix B details the optimized constellations and thresholds for PAM4 and PAM8 for various channel scenarios. Section IV examines the transceiver optimization of PAM IM/DD links with SOA pre-amplification. We compare improvements with standard PAM and other optimized PAM constellations for amplified links. We also address the complexity of the deep learning decoder. We present an experimental validation in Section V for the case of memoryless channels. Finally, Section VI provides concluding remarks.

## II. Preliminaries

### A. Autoencoder structure

In Fig. 1 we show end-to-end learning via an AE, leveraging deep neural networks (DNNs). The encoder will produce the PAM levels, while the decoder will generate symbol decisions. We tailor the neural network (NN) structures to suit the requirements of an IM/DD PAM communications channel.

Random bits are Gray-mapped into PAM symbols. Symbols are mapped to one-hot vectors whose length is equal to the PAM order $M$. The encoder maps symbols to PAM levels $s$ (a constellation) to maximize cross-entropy following the decoder step. The constellation is normalized to unit power, $E[s^2] = 1$. The channel introduces impairments. A memoryless non-Gaussian noise channel will be examined in Section III, whereas a channel with memory will be considered in Section IV.

If the channel is memoryless, we use a scalar input to the decoder. For the case with CD the channel has memory and we input a vector of sequential receiver samples. We found that five sequential inputs are sufficient for the fiber lengths we study. The decoder generates a score vector $\vec{r}$.Symbol decisions are found with the softmax function

$$S(r_i) = \frac{e^{r_i}}{\sum_{j=1}^{M} e^{r_j}}, \tag{1}$$

where $r_i$ represents the $i^{th}$ entry in the scores vector $\vec{r}$.

The DNN weights in the encoder and decoder are adapted to maximize the cross-entropy between the symbol decisions and the true transmitted symbols. The weights are updated through the back-propagation algorithm. Upon model convergence, we generate random test data to estimate the bit error rate (BER) performance. The hyper-parameters of the AE used in the rest of this paper are presented in Table I. The DNN structure and parameters are chosen to keep complexity low, as is appropriate for an IM/DD PAM system.

TABLE I
Hyper parameters of the AE

| Hyper parameter | Value |
|---|---|
| hidden layers of encoder/decoder | 2 |
| nodes per encoder/decoder layer | 15 |
| learning rate | 0.05 |
| activation function | CELU |
| optimizer | Adam |
| batch size | 2048 |
| loss function | CrossEntropyLoss |
| decoder input sequence size (having memory) | 5 |
| decoder input sequence size (no memory) | 1 |

### B. Simple mixed noise memoryless channel

All optical systems are subject to thermal noise during photodetection. For amplified systems, the ASE produced during amplification often dominates thermal noise. The photodetection process squares the ASE noise and adds the thermal noise, as illustrated in Fig. 2. Although both the thermal and ASE noise follow Gaussian distributions, the square law process in IM/DD systems leads to the detected noise being a mixture of Gaussian (thermal) and chi-square (photodetected ASE) noise.

We define $\sigma^2_{ASE}$ as the ASE noise variance, and $\sigma^2_{th}$ as the thermal noise variance. As shown in appendix A, the total noise variance is given by

$$\sigma^2_N = 4E[s^2]\sigma^2_{ASE} + 3\sigma^4_{ASE} + \sigma^2_{th},$$

where $s$ is the PAM signal. To assess the relative influence of ASE and thermal noise, we introduce $\alpha$ as the percentage of total noise attributed to thermal noise that is,

$$\alpha\sigma^2_N = \sigma^2_{th}, \tag{2}$$

and

$$(1-\alpha)\sigma^2_N = 4\sigma^2_{ASE} + 3\sigma^4_{ASE}, \tag{3}$$

where we assume unit signal launch power ($E[s^2] = 1$).

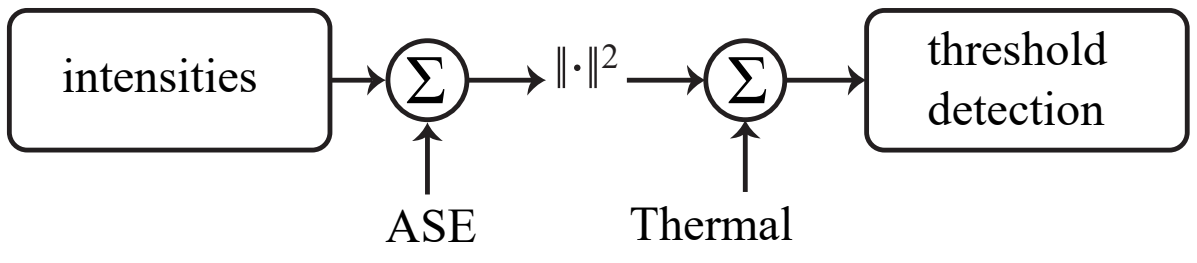

Fig. 2. Memoryless channel model for mixed noise types

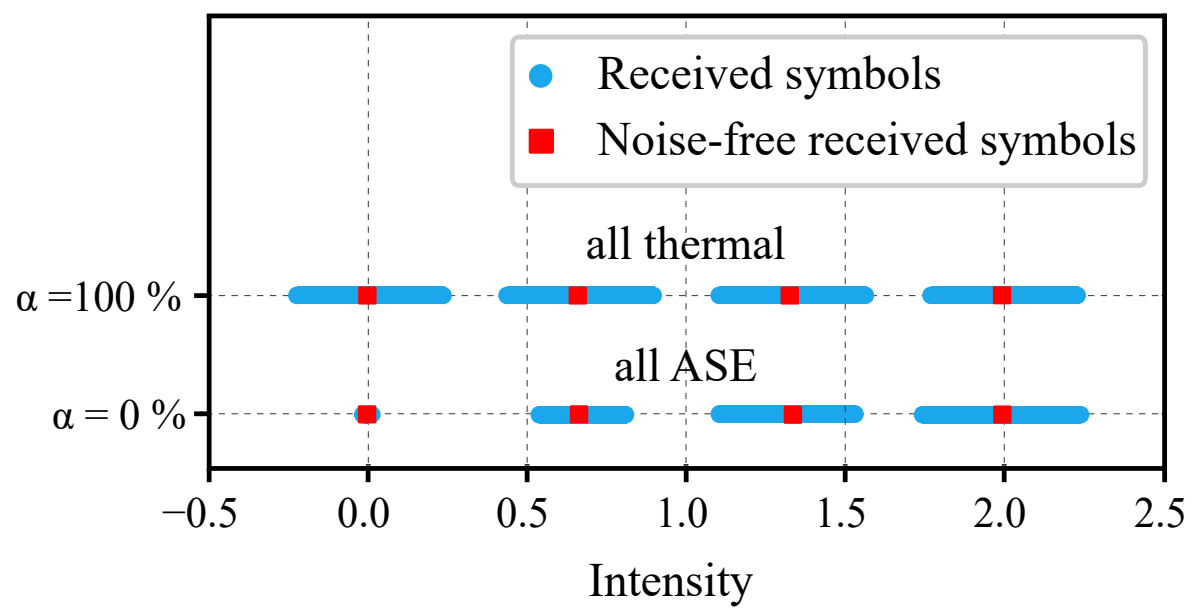

Fig. 3. Received symbols for standard PAM4 for noise that is zero (square markers), all thermal, or all ASE.

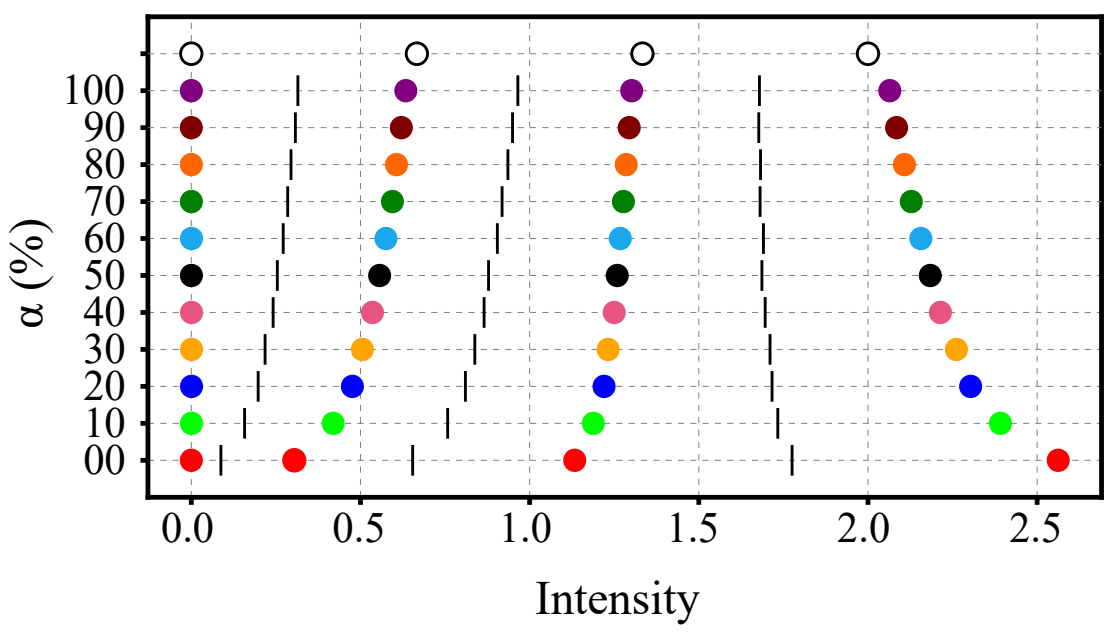

Fig. 4. AE-optimized PAM4 constellations at 18 dB SNR for various α (circles) and their optimized thresholds (vertical lines).

## III. Case without CD

We plot in Fig. 3 the received standard PAM4 symbols in the presence of all thermal noise (upper scatter plot) or all ASE noise (lower scatter plot). For all thermal noise, the standard PAM4 constellation achieves the lowest BER. For all ASE, there is less noise at smaller amplitudes, making tighter spacing at low amplitude desirable. For cases mixing ASE and thermal noise as parameterized by α, we use AE to find the optimal PAM constellation.

For a memoryless channel, one-shot detection is optimal. Therefore, we use AE to find constellations and detection thresholds. Once identified, the AE encoder and decoder are discarded. Consequently, the complexity of the standard and the AE-optimized PAM system is the same.

### A. Optimized constellations

Using the simple channel in Fig. 2, we fix $\sigma_N^2$ and α. We use the AE in Fig. 1 to find PAM4 and PAM8 constellations that maximize cross-entropy, or equivalently minimize BER. For each PAM4 and PAM8 constellation we find the SNR given by $\mathrm{E}[s^4]/\sigma_N^2$. Recall that we normalize to $\mathrm{E}[s^2] = 1$, however, $\mathrm{E}[s^4]$ will vary with constellations $[P_0, P_1, ..., P_{M-1}]$.

In Fig. 4, we give optimized PAM4 constellations with filled circle markers for each α at a SNR of 18 dB. The empty circle markers at the top of the figure are the standard PAM4 constellation. For the case of all thermal noise (α = 100%), the AE constellation is very similar to standard PAM4, i.e., equally spaced levels. As α decreases, reflecting scenarios with increasing ASE noise dominance, the constellation points are tightly spaced at lower amplitudes to take advantage of the reduced noise present there.

The constellations should be optimized for each SNR and each α. The optimizied constellations are given in Appendix B. The parameter α has the greatest impact on symbol spacing. In general, the highest intensity level varies the most with SNR.

### B. Optimized thresholds

As this idealized channel is memoryless, one-shot threshold detection is optimal. Midpoint thresholds are no longer optimal for mixed noise. We use the bisection method to determine the optimal thresholds using the AE-decoder. For any two adjacent constellation points, we initialize the threshold to their midpoint. Points 1e− 5 on either side of the candidate threshold are sent to the decoder. If both points are classified to the same symbol, the threshold is shifted in the opposite direction of the detected symbol. This procedure continues until the two test points are classified to distinct symbols.

The same algorithm is iteratively applied to identify thresholds between other adjacent constellation points. In Fig. 4, we give optimized PAM4 thresholds with vertical lines for each α at a SNR of 18 dB. The threshold is close to the midpoint for large α, that is, nearly Gaussian statistics. In Appendix B, we present the AE-optimized PAM4 and PAM8 detection thresholds over various SNR and α.

### C. Performance

We used Monte Carlo trials to estimate the BER of the system in Fig. 2 using optimized PAM4 and PAM8 constellations and thresholds reported in Appendix B. This modulation/detection has the same complexity as standard PAM. The AE was only used to find constellations and thresholds. We count at least 100 errors when estimating the BER, transmitting 500,000 symbols. We note in passing that PAM4 and PAM8 have a Grey mapping, so that BER is the symbol error rate (SER) divided by the number of bits per symbol.

Figure 5a plots the BER vs. SNR. Results for PAM8 are in solid lines, and PAM4 in dashed lines. Each curve is for fixed α; from left to right, α increases by 10% at each curve. We clearly see the improved BER when optimized constellations and thresholds can take advantage of the non-Gaussian noise statistics at low α.

To quantify the improvement of AE-optimized vis-à-vis standard PAM , we examine the required SNR at BER of $3.8e^{-3}$. Figure 5b gives the required SNR versus α under PAM4 and PAM8 for both standard and AE-optimized transceivers. Standard PAM sees a 1 dB degradation for all-ASE (α = 0). Optimized PAM sees nearly a 4-dB improvement. For all thermal noise (α = 100%) the solutions give the same result.

## IV. Case with CD

The previous section laid the groundwork for AE optimization for ideal, memoryless systems. In this section, we address channel memory in the form of accumulated CD. We examine systems using a SOA for pre-amplification, and thus introducing ASE and producing the noise mix examined in the previous section.

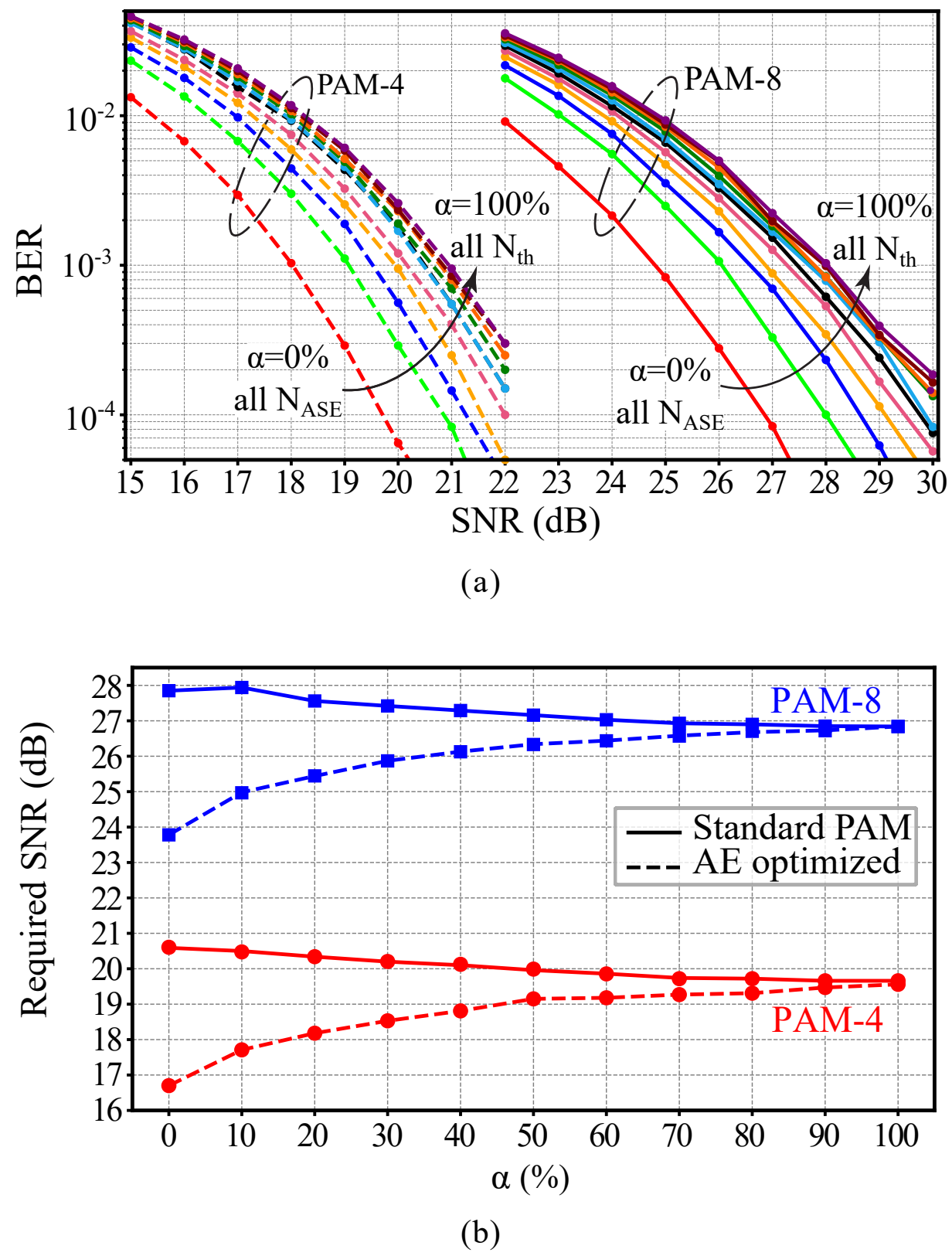

Fig. 5. (a) BER vs. SNR of AE-optimized PAM4 and PAM8 transceivers for various α, and (b) required SNR at BER=3.8e$^{-3}$ vs. α for standard and AE-optimized PAM4 and PAM8.

## A. System model

Figure 6 shows the block diagram of an IM/DD PAM system with non-zero dispersion and SOA pre-amplification. We assume 53 Gbaud transmission using Nyquist signaling to minimize occupied bandwidth and avoid ISI. The otherwise ideal modulator has a 10 dB extinction ratio. Transmission over standard single mode fiber (SMF) will see CD that varies with wavelength. The receiver thermal noise is fixed at -73 dBm. Without CD or amplification, this system requires -13 dBm received optical power to achieve a BER of 1.8e-4 for PAM4.

The SOA pre-amplifier has a noise figure of 6 dB and gain from 0 to 20 dB. Greater SOA gain reduces required optical power for the target BER, often termed sensitivity improvement. Adjusting SOA gain changes the split between thermal/shot noise and ASE noise; that is, there is an equivalence between α and SOA gain.

We simulate a four-wavelength, coarse wavelength division multiplexing (WDM) system in the 1300 nm band. For the three wavelengths with non-zero dispersion, the system has memory. We use the sequential input version of the decoder in Fig. 1 to find the AE-optimized constellations.

Standard PAM uses a linear equalizer to mitigate CD, followed by threshold detection. As threshold detection is suboptimal in a channel with memory, we will retain the AE decoder to take advantage of its enhanced capability to combat CD. This will increase complexity (see section IV-E), but we have taken pains to keep computations low.

## B. Previous iterative optimization technique

In [3], constellations and thresholds were optimized for a similar SOA amplified IM/DD PAM system. They used an iterative, numerical approach to find the constellation points sequentially. We will compare the performance of our AE-optimized constellation and AE decoder with their constellation and threshold approach.

The first PAM level is transmitted at the lowest intensity permitted by the modulator extinction ratio. The decision threshold is found for a target SER (3.6e-4 in our simulations) using the probability distribution found by 1) a theoretical approximation at zero dispersion, or 2) Monte Carlo simulations (or experimental measurements) that can include CD effects. They sweep the next level of the PAM symbol to find one attaining the SER target when using the previously selected lower threshold. It is a sequential approach: as each symbol level/threshold is fixed, the next is found. At the end of this process, all PAM levels and decision thresholds are identified.

As an illustration, we find constellations and thresholds at 53 Gbaud with 20 dB of SOA gain, that is, smallest α and the most non-Gaussian noise. We present received histograms for three scenarios. The first scenario has no dispersion, and Fig. 7a gives a histogram of the sampled received signal. Orange bars show the decision thresholds, unchanged for the three scenarios. The second scenario has accumulated dispersion from 1 km transmission at 1250 nm, with histogram in Fig. 7b. In Fig. 7c we use a linear equalizer, then plot histograms. Clearly the histograms are different in each scenario, although the decision thresholds remain unchanged per this method. The iterative approach precludes the introduction of memory or equalization to assist in finding levels or thresholds.

## C. Performance comparison at 1291 nm

We simulate a coarse WDM O-band system with signals at 1271 nm, 1291 nm, 1310 nm, and 1331 nm with dispersion of -3.74, -1.79, 0, and 1.92 ps/nm-km, respectively. At the zero-dispersion 1310 nm wavelength, we expect similar performance for both optimization techniques. This was borne out in our simulations, with similar constellations, thresholds, and performance for the two optimizations. We focus on 1291 nm in this section, and add 1271 nm and 1331 nm in the next section.

To achieve BER of 1.8e-4 at zero fiber length and zero SOA gain requires -13 dBm received power. We define sensitivity improvement as the reduction in required received power for the same BER. For a given fiber length and SOA gain, we run Monte Carlo simulations of PAM4. From the BER vs. received power curves, we extract the sensitivity improvement.

Figure 8a presents sensitivity improvement vs. SOA gain for three systems with 1 km of accumulated dispersion. The circle markers are for standard PAM4 with MMSE equalization and midpoint thresholds. The diamond markers are for iterative PAM4 with MMSE equalization and optimized thresholds. The square markers are for AE-optimized PAM4 and an AE decoder. An inset shows the PAM4 levels for the iterative and the 20 dB SOA gain AE-optimized cases.

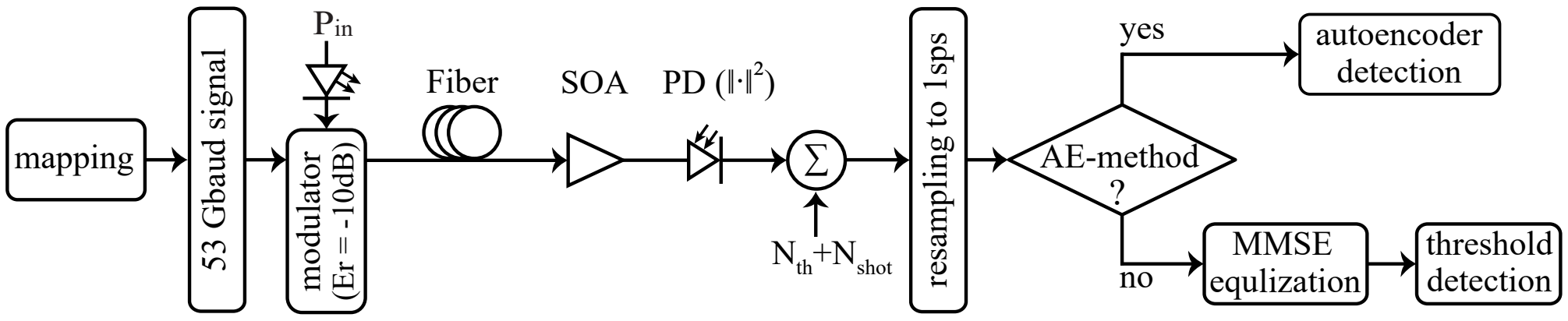

Fig. 6. Block diagram of the amplified IM/DD PAM system with non-zero dispersion

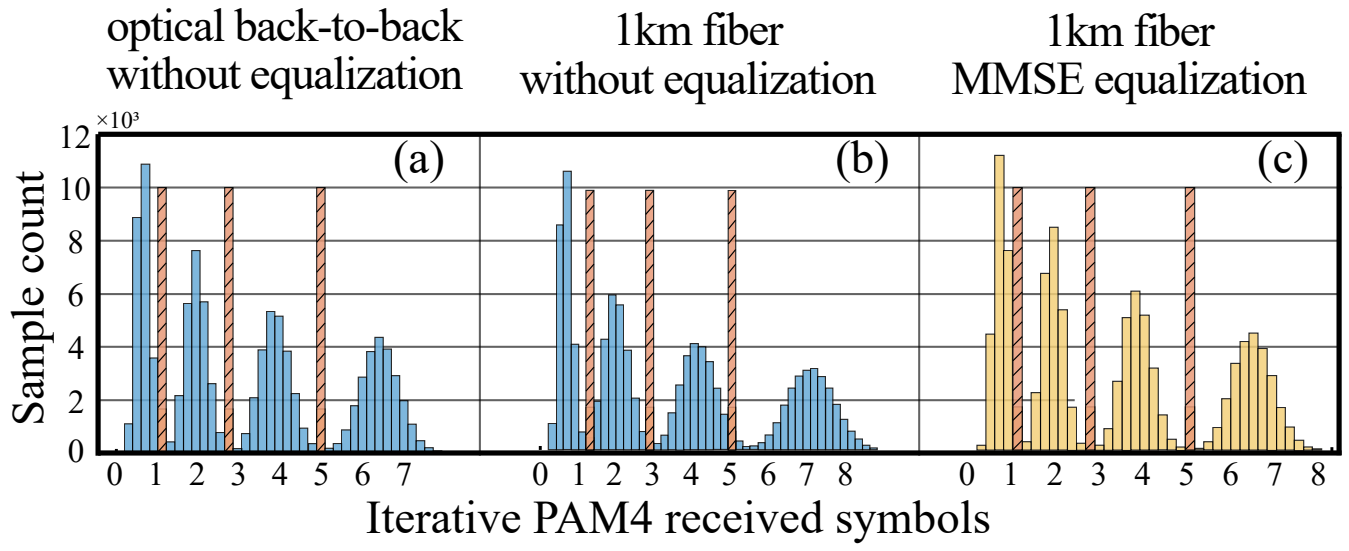

Fig. 7. Three scenarios for a constellation and thresholds found by the iterative method. Received histograms for a) no dispersion, and for 1 km of dispersion at 1250 nm with b) no equalizer, and c) a MMSE equalizer.

To facilitate comparison with the results in the previous section, for each SOA gain we associate an equivalent α, where α is the percentage of total noise power attributable to the fixed -73 dBm thermal noise power. We observe that at this distance the iterative and AE solutions have similar performance. At high SOA gain, both offer substantial improvement (over 2 dB) compared to standard PAM.

Results in Fig. 8b are for 3 km of accumulated dispersion. The iterative solution is now only slightly better than standard PAM, while the AE solution sees even greater improvement (over 3 dB). This trend continues as the length of the fiber increases. This can be directly attributed to the constellation and thresholds in the iterative method being optimized for zero dispersion. The most significant performance gains vis-à-vis standard PAM are observed at low α values, i.e., where SOA amplification is high. This is consistent with our analysis in Section III.

## D. Performance comparison at three wavelengths

In Fig. 9 we plot the relative performance improvement for AE vs. SOA gain for each of the three non-zero dispersion wavelengths. One group of curves (dashed) has AE improvement compared to standard PAM, the other group (solid) has AE improvement compared to the iterative method. Markers are used to indicate wavelengths. Figure 9a is for 1 km of accumulated dispersion, while Fig. 9b is for 3 km.

Similar trends are observed for the three non-zero dispersion wavelengths. The improvement of AE over standard PAM transceiver grows with the SOA gain, but remains roughly stable over distance. The iterative method breaks down with accumulated dispersion. The AE decoder uses nonlinear behavior to combat CD more effectively than MMSE equalization.

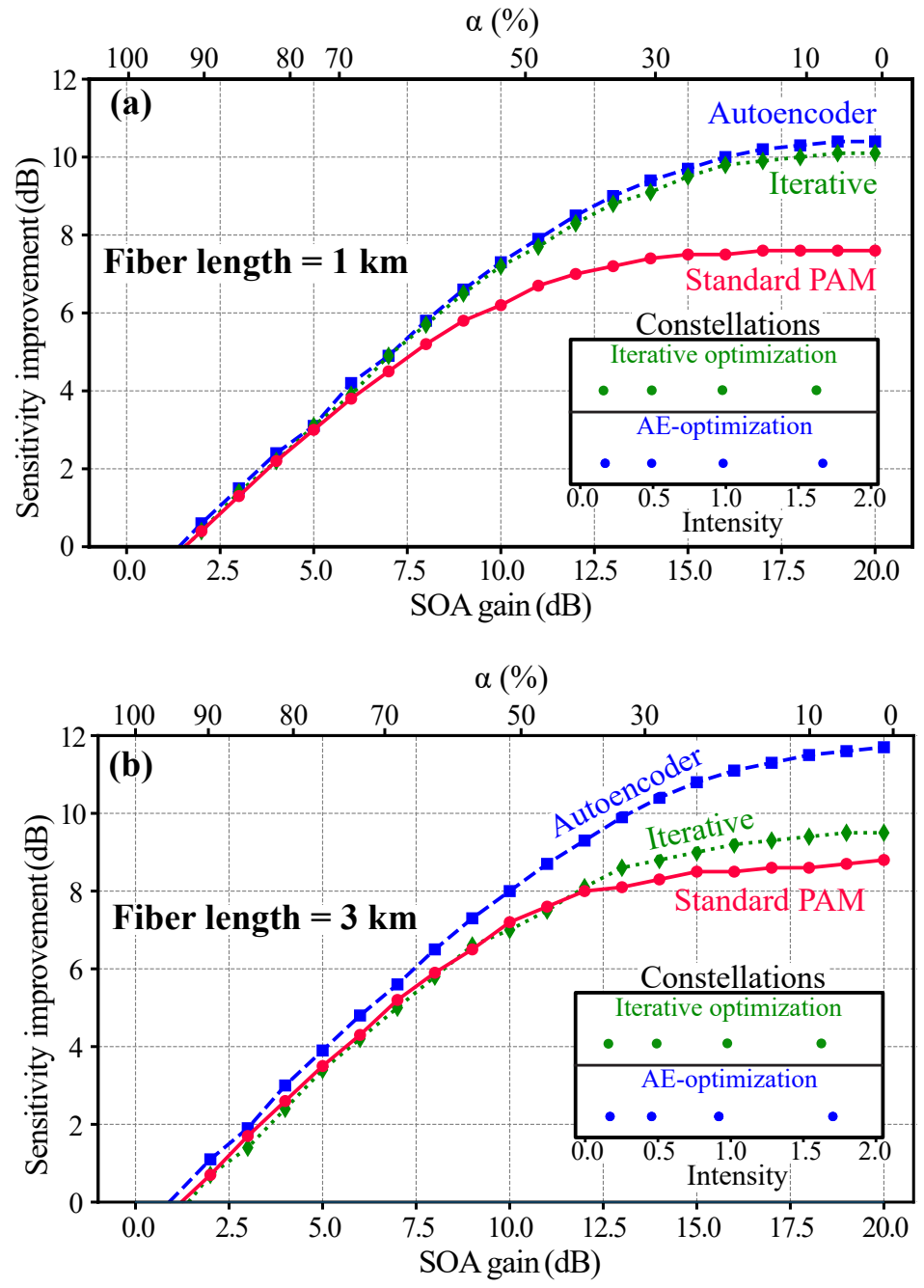

Fig. 8. Sensitivity improvement vs. SOA gain for three PAM4 systems at 1291 nm and fiber length of a) 1 km, and b) 3 km; upper x-axis gives percentage of receiver noise attributable to thermal noise.

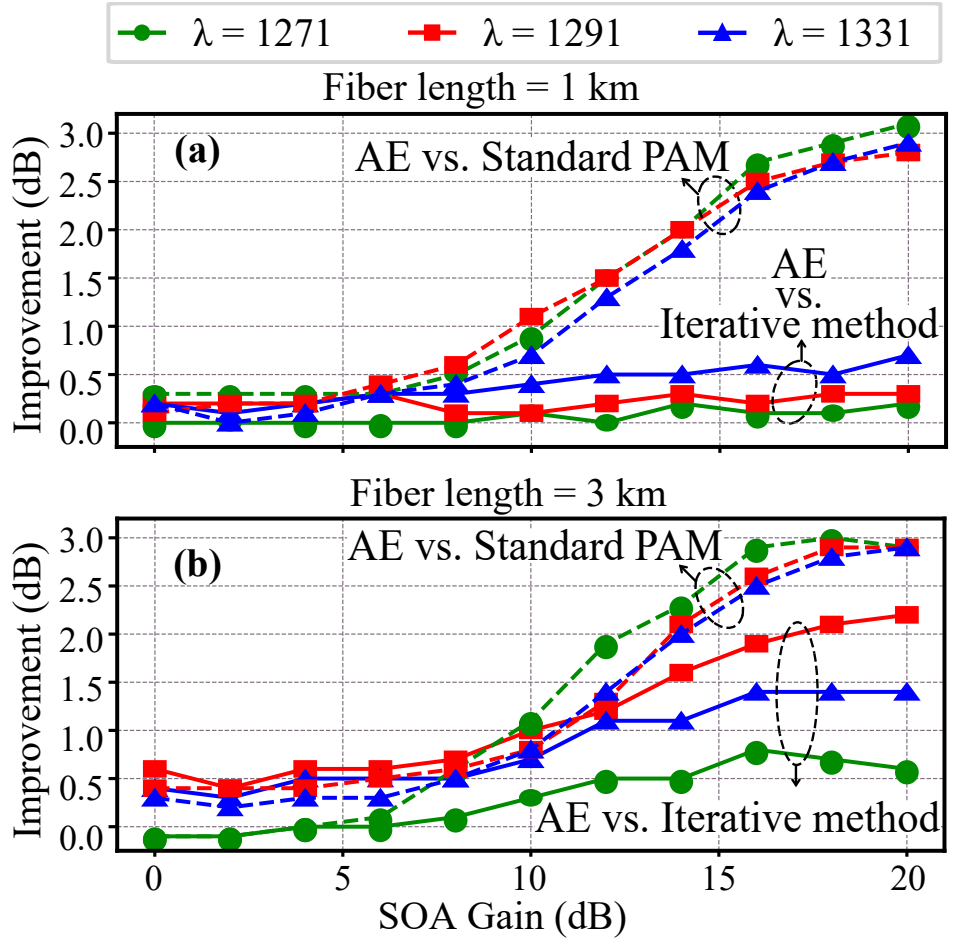

Fig. 9. Improvement obtained with AE vs. standard PAM (dashed) or the iterative method (solid) at 1271 nm, 1291 nm, and 1331 nm, for (a) 1 km, and (b) 3 km.

At 53 Gbaud, both standard and AE constellations have reliable communications beyond 3 km, while the iterative constellation/thresholds are completely unrecoverable. We fix the SOA gain to 20 dB. With the AE constellation we examine the reach of each of the three non-zero dispersion wavelengths.

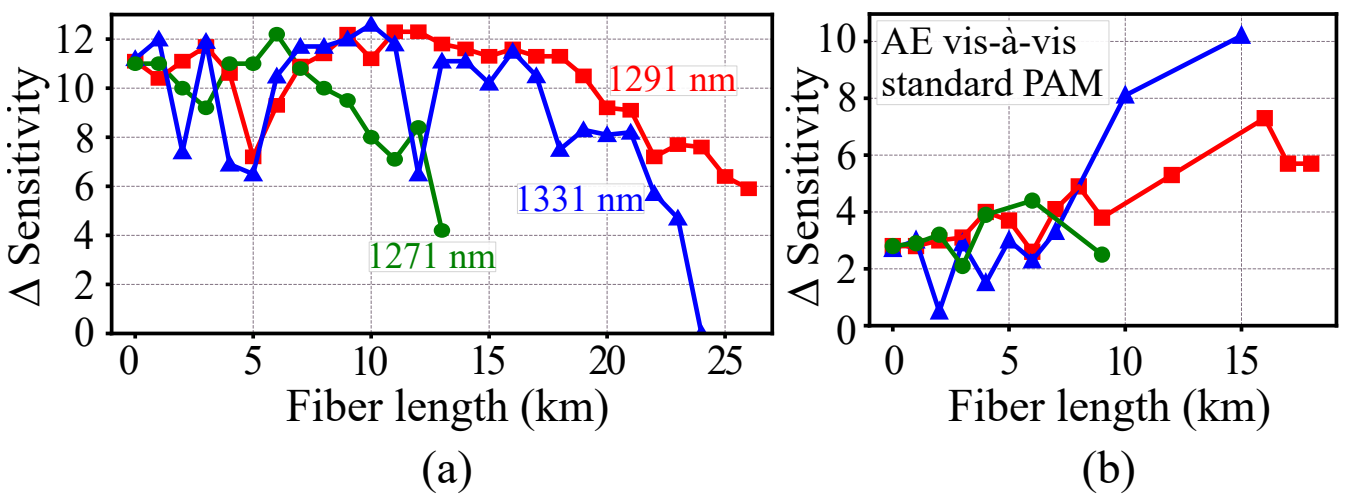

Fig. 10. For SOA gain of 20 dB, sensitivity improvement vs. fiber length at three wavelengths for a) AE-optimized PAM4 compared to no SOA, and b) AE-optimized PAM4 compared to standard PAM; iterative constellations were only recoverable up to 3 km.

At each fiber length, we find the AE constellation performance with and without the 20 dB SOA gain. In Fig. 10a we plot the sensitivity improvement vs. fiber length for the AE constellation.

As long as the sensitivity improvement remains above 4 dB, we can respect the KP4 forward error correction (FEC) threshold. This leads to reach of 13 km, 26 km, and 23 km at 1271 nm, 1291 nm, and 1331 nm, respectively.

The standard constellation can also support longer fiber distances, although only up to 9 km, 18 km, and 15 km at 1271 nm, 1291 nm, and 1331 nm, respectively. This corresponds to increased reach of as much 8 km (1331 nm). In Fig. 10b we plot the sensitivity improvement for the AE constellation compared to the standard constellation. At fiber lengths below 10 km, the AE enjoys on overage a sensitivity advantage of 2 dB. Beyond 10 km the advantage increases substantially for 1291 nm and 1331 nm.

## E. Complexity Analysis

We use the AE encoder to find optimized PAM levels, but then abandon the encoder. Hence, all additional complexity derives from the use of the AE decoder in channels with memory.

Per Tab. I, a sliding window of length 5 is acted upon by 15 hidden units in each of two layers. Therefore, the first hidden layer involves (5 × 15) multiplications. The second hidden layer involves (15 × 4) multiplications to output 4 scores for PAM4.

The decoder has a continuous exponential linear unit (CELU) activation function; we use

$$\mathrm{CELU}(x) = \max(0, x) + \min(0, e^{x} - 1). \tag{4}$$

Both the CELU and the softmax in (1) are applied to a vector with length equal to the PAM order $M$. The softmax and CELU can be approximated by a third-degree Taylor series expansion, that is, two multiplications. As we normalize to the range [-1, 1], this truncated Taylor series offers good precision [22]. The final number of multiplications is therefore on the order of

$$(5 \times 15) + (15 \,□ M) + M \,□ 2 \,□ 2 \tag{5}$$

or 151 for PAM4 and 227 for PAM8.

Conventional approaches use a MMSE linear filtering and threshold detection. For the systems we examined, we saw no improvement in BER when using more than 15 taps in our simulations (dispersion in coarse WDM in O-band) and 19 taps in our experiments (ISI from non-ideal components). Although our AE decoder has one order of magnitude more multiplications, the added complexity is not excessive. In contrast, previously proposed end-to-end learning models can employ multiple hidden layers, each with tens or hundreds of neurons, in both the encoder and decoder NNs [17]–[20].

## V. Experimental Validation

We experimentally validated the improvement of AE-optimized constellations in the mixed noise case of Section III with high ASE. We examined PAM4 in an optical system with a silicon photonics external modulator. The modulator uses vertical grating couplers with □10 dB loss back-to-back requiring several stages of amplification, which leads to high ASE.

### A. Setup

The experimental setup is shown in Fig. 11. The silicon Mach-Zehnder modulator (MZM) is designed to operate in the C-band. It has 4-mm phase shifters implemented as lateral PN-junctions on both arms for single drive push-pull operation. The MZM has traveling-wave electrodes, providing a 40 GHz electro-optical bandwidth and a $V_\pi$ of around 8 V, by carrier depletion [23]. The length difference between each arm is set to 100 µm, leading to a sinusoidal transmission function at the MZM output.

We use heaters on both arms to bias at a quadrature point, while keeping a low reverse bias voltage on the electrodes to maximize the modulation efficiency. Since the MZM voltage response has a negative slope around the chosen quadrature point, the data to the digital-to-analog convert (DAC) was inverted. We tuned a Mach-Zehnder interferometer coupler before the MZM to adjust the coupling ratio between modulator arms and maximize the modulation depth.

The maximum achievable OSNR when operating near the quadrature point is 30 dB in our setup. To sweep OSNR, we load the received signal with additional ASE.

We generate a random sequence of $2^{17}$ PAM4 symbols for either standard PAM4 or the AE-optimized constellation. We synthesize the 60 Gbaud radio frequency signal with a DAC at 100 GSa/s, amplify to 4 V peak-to-peak and send it to the MZM. The input optical power is boosted to 24 dBm by an erbium-doped fiber amplifiers (EDFA) to overcome the insertion loss of the chip and to introduce ASE noise. At the receiver side, the optical signal is pre-amplified using another EDFA, increasing the sensitivity, and adding more ASE noise. We use a separate EDFA (not shown) as a noise source to tune the received OSNR from 18 dB to 30 dB. The optical signal is converted to the electrical domain via a 70-GHz bandwidth photodiode and sampled with a 65 GHz real-time oscilloscope at 160 GSa/s.

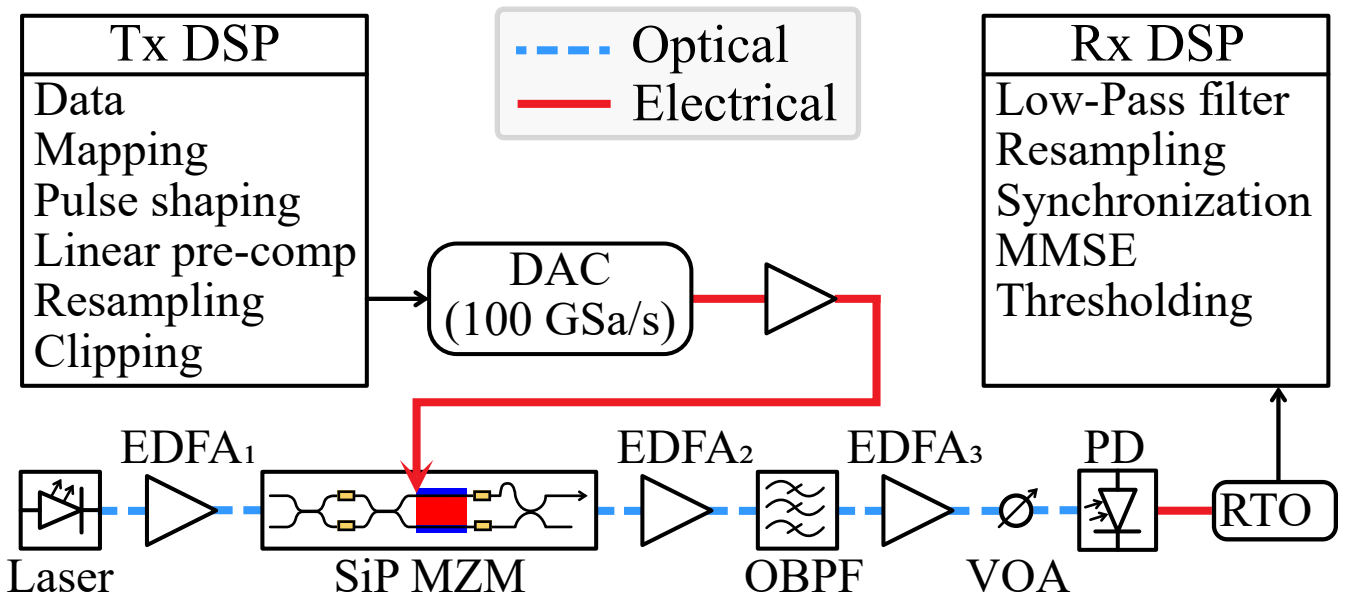

Fig. 11. Experimental setup

For both PAM4 constellations we use the transmitter and receiver digital signal processing (DSP) in Fig. 11. The symbols are shaped with a root-raised cosine pulse with a roll-off factor of 0.1, with a finite impulse response (FIR) filter to compensate for the frequency response of the DAC and radio frequency (RF) driver. At the receiver side, we low pass filter to remove out-of-band noise and resample to 2 samples per symbol. We find MMSE filter taps using known transmitted data. We tried several time-domain filter lengths and settled on 19 taps. More taps did not provide a significant increase in performance for the standard constellation.

## B. Results

Experimental demonstrations include nonideal effects from various components. The components in this setup have bandwidths well beyond the Nyquist signal bandwidth of about 30 GHz, so the channel is not band-limited. However, the wideband photodetector has no transimpedance amplifier, so thermal noise is non-negligible. We use several stages of amplifications, so ASE will be prominent and thermal noise will not dominate. Therefore, we tested several constellations to identify the one that yielded the best performance. All AE experimental results are for the constellation of (0 0.98 1.39 1.75), which can be written as received symbols of (0, 0.97, 1.92, 3.06) when normalized to the same power as received symbols (0, 1, 2, 3).

The BER is plotted versus OSNR in Fig. 12a. We note that the results originally reported in [21] had an error in the OSNR axis that is corrected here. We see that the optimized constellation always outperforms standard PAM. At BER of $10^{-4}$ there is over 4 dB improvement. We used experimental data to train the AE decoder. After convergence, we extracted the experimental optimal thresholds and discarded the decoder. The BER was found using threshold detection.

Knowledge of the transmitted random sequence (no pseudo-random sequences were used) allowed us to plot the empirical conditional probability densities at each PAM level. We chose data collected with similar BER 3e-3 for both constellations (OSNR 22 and 24 dB), and plotted histograms to the left in Fig. 12b. We also examined the histograms for OSNR of 30 dB for both constellations and plotted them on the right side in Fig. 12b. See appendix C for the normalization used.

We first note that in all histograms we see that the conditional probabilities are distinct for each symbol. The variance of logical zero is much smaller than that of logical 3. As expected, noise on the highest power PAM level has a significantly longer tail. For each symbol, we provide the transmitted coordinates (in units proportional to received power), as well as the experimental received means. Some distortion is introduced, causing logical 1 and 3 to decrease in value and logical 2 to increase in value. This is experienced by both constellations.

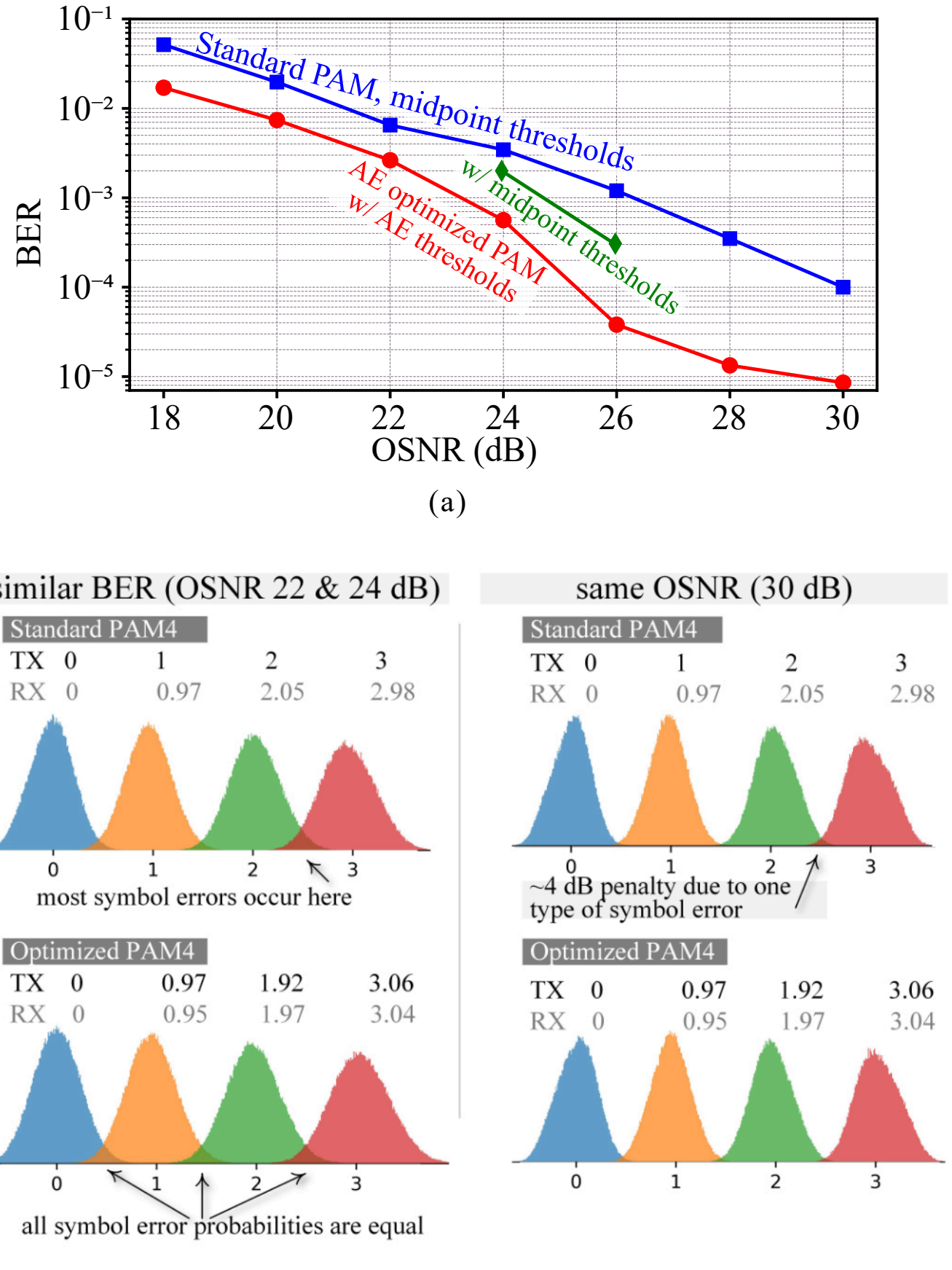

Fig. 12. (a) Bit error rate vs. OSNR and (b) received conditional probabilities (empirical).

The overlap in the conditional probabilities corresponds to symbol errors. In the case of standard PAM the symbol errors between logical 2 and 3 dominate. In the AE optimized case, all error types have similar overlap, thus minimizing total errors. Means for logical 2 and 3 are similar for both constellations (2.05, 2.98) vs. (1.97, 3.04). However, on the right of Fig. 12b (same received power), even these small differences can lead to an improvement of over 4 dB at OSNR of 30 dB.

At OSNR of 24 and 26 dB we passed the received data through a threshold detector set to the midpoints of the transmitted AE symbols. The BER at 24 dB deteriorated by about an order of magnitude from 3e-4 to 2e-3 when using midpoint thresholds rather than thresholds found by the AE decoder. Although the thresholds were similar (midpoints $[0.485, 1.45, 2.48]$ vs. optimized $[0.49, 1.48, 2.51]$ at 26 dB), the disparate widths of the conditional probabilities per symbol lead to the difference in BER.

## VI. Conclusion

We examined the advantages of an AE-optimized transceiver for PAM in amplified IM/DD links. We demonstrated the potential of combining optimized constellation and nonlinear NN decoder in enhancing IM/DD throughput and reach. For a simple memoryless channel, performance was improved by 4 dB with no added complexity. For SOA-amplified O-band WDM links with non-zero dispersion, reach extension of 4-8 km was shown via simulation at minor complexity increase over linear filtering. For an amplified C-band link with high ASE, we achieved 4 dB SER improvement experimentally.

The successful adaptation of our method to various channel conditions underscores its versatility and robustness, and the potential of end-to-end learning to transform optical communication systems.

## Acknowledgments

This project is funded by NSERC (IRCPJ 546377 - 18).

## Appendix A
## Noise variance calculation

We employ a straightforward analytical model to characterize the mixture of different types of noise. The received signal before the photodiode is $s + N_{ASE}$, where s is the transmitted PAM signal with intensity levels of $[P_0, P_1, .., P_{M-1}]$. The received signal after photodetection is given by

$$(s + N_{ASE})^2 + N_{th}, \tag{6}$$

where $N_{th}$ is the thermal additive white Gaussian noise (AWGN) with zero mean and variance $\sigma_{th}^2$. Given $N_{ASE}$ is a zero-mean Gaussian random variable with variance $\sigma_{ASE}^2$ for optical ASE, the photodetected electrical ASE noise follows a non-central chi-square distribution with one degree of freedom [24].

The detected power attributable to the transmitted signal is given by

$$I^2 = E[(Rs^2)^2] = R^2 E[s^4] = E[s^4], \tag{7}$$

where R is the photodiode responsivity that is assumed to be 1 A/W. Again using $R = 1$, the detected power attributable to noise and signal-noise beating is

$$\sigma_N^2 = E[(2sN_{ASE} + N_{ASE}^2 + N_{th})^2],$$

where the first variance term is for Gaussian but signal-dependent noise, the second variance term is for chi-square noise, and the third variance term is for Gaussian noise. Using the independence of signal and noise we have

$$\begin{aligned}\sigma_N^2 = E\big[&4s^2N_{ASE}^2 + 4sN_{ASE}^3 + N_{ASE}^4 + N_{th}^2 \\ &+4sN_{ASE}N_{th} + 2N_{ASE}^2N_{th}\big] \\ = {}&4E[s^2]E[N_{ASE}^2] + 4E[s]E[N_{ASE}^3] + E[N_{ASE}^4] \\ &+ E[N_{th}^2] + 4E[s]E[N_{ASE}]E[N_{th}] \\ &+ 2E[N_{ASE}^2]E[N_{th}],\end{aligned}$$

Using the properties of the standard normal for ASE and thermal noise,

$$\begin{aligned}\sigma_N^2 = {}&4E[s^2]E[N_{ASE}^2] + 4E[s]\underbrace{E[N_{ASE}^3]}_{0} + E[N_{ASE}^4] \\ &+ E[N_{th}^2] + 4E[s]\underbrace{E[N_{ASE}]}_{0}\underbrace{E[N_{th}]}_{0} \\ &+ 2E[N_{ASE}^2]\underbrace{E[N_{th}]}_{0} \\ = {}&4E[s^2]E[N_{ASE}^2] + E[N_{ASE}^4] + E[N_{th}^2],\end{aligned}$$

As the standard normal variable has a fourth moment equal to three times the square of its variance [24] (5-75), we have

$$\begin{aligned}\sigma_N^2 &= 4E[s^2]\sigma_{ASE}^2 + 3(\sigma_{ASE}^2)^2 + \sigma_{th}^2 \\ &= 4E[s^2]\sigma_{ASE}^2 + 3\sigma_{ASE}^4 + \sigma_{th}^2.\end{aligned}$$

## Appendix B
## Constellations and thresholds for different SNR values

Fig. 13 shows the constellations and thresholds for various SNR and $\alpha$ values. For each $\alpha$, eight distinct SNR scenarios are examined, distributed between ASE noise and thermal noise according to $\alpha$. The SNR range is from 22 to 29 dB for PAM8 and from 15 to 22 dB for PAM4. The dependence of both constellation and thresholds on SNR is low.

## Appendix C
## Histogram normalization

The signals generated with the DAC had the same power for all the tested constellations (normalized constellations and the same peak-to-peak voltage). The received power was, of course, always positive. To easily compare quantities, in the experimental Section V we give histograms with an x-axis of logical 0, 1, 2 and 3. The actual received powers r were manipulated as follows to produce the histogram values h

$$h = (r - \mu_0) \cdot \sqrt{14} \div \sqrt{\sum_{i=0}^{3} (\mu_i - \mu_0)^2},$$

where $[\mu_0, \mu_1, \mu_2, \mu_3]$ are the mean received powers of the symbols in ascending power.

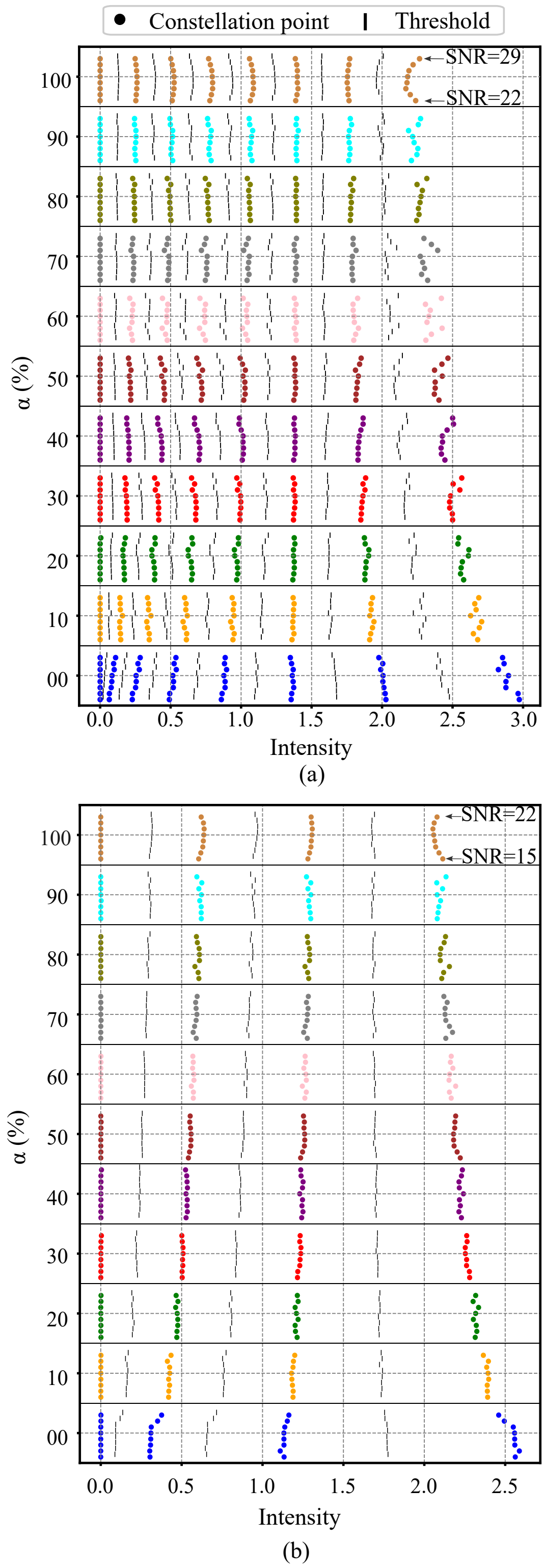

Fig. 13. Constellations and related thresholds for different SNRs for (a) PAM8 and (b) PAM4.